\documentclass[a4paper,11pt]{article}
\pdfoutput=1 

\usepackage{jcappub} 

\usepackage[T1]{fontenc} 
\usepackage{graphicx}
\usepackage{bm}
\usepackage{amsmath}
\usepackage{amssymb}
\usepackage{graphicx}
\usepackage{xcolor}
\usepackage{mathtools}
\usepackage{cancel}
\usepackage{changes}
\usepackage{adjustbox}
\usepackage{threeparttable}
\usepackage[english]{babel}
\usepackage{ragged2e}
\usepackage{blindtext}

\makeatletter
\gdef\@fpheader{}
\makeatother

\def\lsim{\mathrel{\rlap{\lower4pt\hbox{\hskip1pt$\sim$}} \raise1pt\hbox{$<$}}}
\def\gsim{\mathrel{\rlap{\lower4pt\hbox{\hskip1pt$\sim$}} \raise1pt\hbox{$>$}}}

\newcommand{\sml}[1]{\textcolor{black}{ #1}}

\begin{document}

\begin{flushright}
   CERN-TH-2022-117 \\
   KIAS-P22049 \\
   APCTP-Pre2022-015
\end{flushright}

\title{\boldmath \huge Cosmic Birefringence by Dark Photon }

\author[a,b]{\Large Sung Mook Lee}
\author[c,d]{\Large Dong Woo Kang}
\author[e,f]{\Large Jinn-Ouk Gong}
\author[e,g]{\Large Donghui Jeong}
\author[a,h,i]{\Large Dong-Won Jung}
\author[a,c]{\Large Seong Chan Park}

\affiliation[a]{\small Department of Physics \& IPAP \& Lab for Dark Universe, Yonsei University, Seoul 03722, Korea}
\affiliation[b]{\small Department of Physics, Korea Advanced Institute of Science and Technology, Daejeon
34141, Korea}
\affiliation[c]{\small School of Physics, Korea Institute for Advanced Study, Seoul, 02455, Korea}
\affiliation[d]{\small Theoretical Physics Department, CERN, CH-1211 Gen\`eve 23, Switzerland}
\affiliation[e]{\small Department of Science Education, Ewha Womans University, Seoul 03760, Korea}
\affiliation[f]{\small Asia Pacific Center for Theoretical Physics, Pohang 37673, Korea}
\affiliation[g]{\small Department of Astronomy and Astrophysics and Institute for Gravitation and the Cosmos, The Pennsylvania State University, University Park, PA 16802, USA}
\affiliation[h]{\small Department of Physics, Chonnam National University, Gwangju 61186, Korea}
\affiliation[i]{\small Cosmology, Gravity, and Astroparticle Physics Group, Center for Theoretical Physics of the Universe, Institute for Basic Science, Daejeon 34126, Korea}

\emailAdd{sungmook.lee@kaist.ac.kr}
\emailAdd{dongwookang@kias.re.kr}
\emailAdd{jgong@ewha.ac.kr}
\emailAdd{djeong@psu.edu}
\emailAdd{dongwon.jung@yonsei.ac.kr}
\emailAdd{sc.park@yonsei.ac.kr}

\abstract{
We study the kinetic mixing between the cosmic microwave background (CMB) photon and the birefringent dark photon. These birefringent dark photon may exist in parity-violating dark sector, for example, through the coupling to axion field. We show that the birefringence of the dark photon propagates to the CMB photon, but the resulting birefringence may not be isotropic over the sky, but will be anisotropic in general.
Moreover, our investigation sheds light on the essential role played by kinetic mixing in the generation of two fundamental characteristics of the CMB: circular polarization and spectral distortion. 
}

\maketitle
\flushbottom

\section{Introduction}

Although Maxwell's theory of electrodynamics upholds parity as a fundamental symmetry, it can be disrupted by introducing a Chern-Simons coupling with a pseudoscalar field $\theta$~\cite{Carroll:1989vb,Carroll:1991zs,Harari:1992ea}:
\begin{align}
{\cal L}_{\rm CS} = g_{\theta} \theta F_{\mu\nu}\tilde{F}^{\mu\nu}\,,
\label{eq:Lcs}
\end{align}
where $g_{\theta}$ is a coupling constant, $F_{\mu\nu}\equiv\partial_\mu A_\nu-\partial_\nu A_\mu$ is the field strength tensor, and $\tilde{F}^{\mu\nu}\equiv2\epsilon^{\mu\nu\alpha\beta}F_{\alpha\beta}/\sqrt{-g}$ is the dual strength with the Levi-Civita symbol $\epsilon^{\mu\nu\alpha\beta}$.
A popular example for the pseudoscalar field is an axion \cite{Weinberg:1977ma,Wilczek:1977pj} or axion-like particles (ALPs) $a(x)$, for which one can identify $g_{\theta} \theta = g_{a} a/f_{a}$, where $g_{a}$ is a coupling constant of the order of fine-structure constant $\alpha=e^2/(4\pi)$, and $f_{a}$ is the axion decay constant. See e.g. \cite{Choi:2020rgn} for a recent review.  
With this interaction, the dispersion relations of two circular polarization modes of the electromagnetic waves differ from each other, i.e. parity is violated. 
The Thomson scattering on the last scattering surface of the cosmic microwave background (CMB) leads to the linear polarization of the CMB~\cite{Rees:1968,Kosowsky:1994cy}.
Thus, the interaction~\eqref{eq:Lcs} yields rotations of the linear polarizations, called ``cosmic birefringence''.
When decomposing the angular distribution of the CMB polarization by the $E$-mode (even-parity) and the $B$-mode (odd-parity)~\cite{Zaldarriaga:1996xe,Kamionkowski:1996ks} the $EB$-cross correlation vanishes in the standard $\Lambda$CDM cosmological model~\cite{Lue:1998mq,Gluscevic:2010vv}.
Therefore, the detection of the $EB$-cross correlations will be a clear smoking gun of parity-violating new physics beyond the standard model of particle physics (BSM). 
Interestingly, recent analyses of the CMB have provided a tantalizing hint of the $EB$-cross correlation that is consistent with an isotropic birefringence signal~\cite{Minami:2020odp,Diego-Palazuelos:2022dsq,Eskilt:2022wav,Eskilt:2022cff} due to the interaction~\eqref{eq:Lcs}.
Since ALPs may contribute to dark matter and/or dark energy, the observation of parity-violating physics in the polarization of the CMB could represent a significant step toward our understanding of the dark sector~\cite{Komatsu:2022nvu}.
In this paper, motivated by the hints of the parity violation in our universe, we investigate the consequences of the parity violation in the dark sector from an alternative interaction that the photon can participate in: kinetic coupling to other massless U(1) gauge fields.
Especially, we assume the new gauge field is completely secluded from the standard model (SM) sector other than the kinetic coupling but is birefringent due to its interactions with dark sector particles.
The paper is organized as follows. 
In Section~\ref{section:model}, we begin by reviewing the model of dark photons 
{with the kinetic mixing to SM and modification of the Maxwell equations.}
In Section~\ref{section:polarization}, we derive the relation of birefringence in SM and that in the dark photon by considering the polarization tensors of each sector.
In Section~\ref{section:implication}, we discuss the implications of our findings and the current and future constraints on the model.
Finally, we conclude in Section.~\ref{section:conclusion}.

\section{Maxwell Equations with Dark Photon Kinetic Mixing}
\label{section:model}

The model consists of the photon of ${\rm U}(1)_{\rm EM}$ denoted by $ \hat{A}^{\mu} $, and a massless dark photon of a dark ${\rm U(1)}_X$ gauge theory denoted by $\hat{A}_X^{\mu}$, whose Lagrangian density contains the following kinetic terms:
\begin{align}
\frac{\mathcal{L}_{\rm kin}}{\sqrt{-g}} 
	 = 
	- \frac{1}{4} \hat{F}^{\mu\nu} \hat{F}_{\mu\nu} - \frac{1}{4} \hat{F}_X^{\mu\nu} \hat{F}_{X \mu\nu} 
	- \frac{\varepsilon}{2}  \hat{F}_{\mu\nu} \hat{F}_X^{\mu\nu} 
    \, ,
\end{align}
where $\varepsilon$ is the kinetic mixing coefficient, $\hat{F}_{\mu\nu} \equiv \partial_{\mu} \hat{A}_{\nu} - \partial_{\nu} \hat{A}_{\mu}$, and $\hat{F}_{X\mu\nu} \equiv \partial_{\mu} \hat{A}_{X\nu} - \partial_{\nu} \hat{A}_{X\mu}$ are the field strength tensors.
The Lagrangian density is conveniently diagonalized with the following linear transformation:
\begin{align}
\binom{\hat{A}^{\mu}}{\hat{A}_X^{\mu}} 
=\begin{pmatrix}
\dfrac{1}{\sqrt{1-\varepsilon^{2}}} & 0 
\vspace{0.3em}\\
-\dfrac{\varepsilon}{\sqrt{1-\varepsilon^{2}}} & 1
\end{pmatrix}
\binom{A^{\mu}}{A_X^{\mu}} 
\, , \label{eq:rotation}
\end{align}
where $ A^{\mu} $ and $A_X^{\mu} $ are, respectively, what we identify as the photon and the dark photon respectively.
The kinetic mixing changes the interaction Lagrangian, {modifying} the interactions of photon and dark photon with the electric and dark-electric currents:
\begin{equation}
    \begin{aligned}
\frac{{\cal L}_{\rm int}}{\sqrt{-g}} \supset 	& ~e j_{\mu} \hat{A}^{\mu} + e_{X} j_{X \mu} \hat{A}_X^{\mu} \approx 
	 \left( e j_{\mu} - \varepsilon e_{X} j_{X \mu}	\right) A^{\mu}
	+e_{X} j_{X \mu} A_X^{\mu} \, ,
	\label{eq:massless interaction}
\end{aligned}
\end{equation}
where we take $\varepsilon \ll 1$.
Note that the photon couples to the dark current $ j_{X \mu}$ with a coupling proportional to the kinetic mixing parameter $ \varepsilon $, but the dark photon is inert to the SM charged matters.
In literature, the coupling between the photon and the dark current is often parameterized as a {\it milli-charge} \cite{Huh:2007zw,Park:2012xq,Fabbrichesi:2020wbt}
\begin{align}
\epsilon  \equiv - \varepsilon \frac{e_X}{e} \, .
\end{align}
Indirect constraints to $\epsilon$ come from the milli-charged particle (MCP) searches. 
The constraints from LEP and LHC allow $\epsilon\lesssim 0.1$ for a MCP mass $\in[6,300]~{\rm GeV}$ \sml{and future $ \Delta N_{\rm eff} $ bound would be able to close this window up to $ \epsilon \lesssim \mathcal{O} (10^{-6})$~\cite{Vogel:2013raa,Fabbrichesi:2020wbt, Berlin:2022hmt,Adshead:2022ovo}}.\footnote{\sml{These constraints are more severe if MCPs consist of all dark matter while we are not assuming this is the case \cite{McDermott:2010pa,Dvorkin:2013cea}. Our following analysis are independent on the constraints of $ \epsilon $ while observational possibilities depends on this.}}
The future experiments such as FerMINI~\cite{Kelly:2018brz} and milliQAN~\cite{Ball:2016zrp} will cover up to $ \epsilon \lesssim \mathcal{O}\left( 10^{-3} \right) $ in this mass range \sml{too}.
For a mass range $m_{\rm MCP} \gsim 1~{\rm TeV}$, there hardly exist constraints on the kinetic mixing coming from MCP.
Also, the total energy density of dark photon is bounded at  the time of CMB (and BBN). Explicitly, we request $\rho_{\gamma_{X}}/\rho_\gamma \leq 0.065$ to be consistent with the CMB bounds on $ \Delta N_{\rm eff}$~\cite{Planck:2018vyg}.
Maxwell's equations for photon and dark photon along with the corresponding currents are
\begin{align}
	\nabla_{\mu}F^{\mu\nu} = 4 \pi \left(j^{\nu} + \epsilon j^{\nu}_{X} \right)
	\quad \text{and} \quad 
	\nabla_{\mu}F_X^{\mu\nu} = 4 \pi j_{X}^{\nu} \, .
 \label{eq:new maxwell}
\end{align}
Both photon and dark photon satisfy the Bianchi identity: $\partial_{\rho} \tilde{F}_{\mu\nu} =\partial_{\rho} \tilde{F}_{X\mu\nu}=0 $. 
In this article, we assume that the dark photon is birefringent which happens when the dark current $j_X$ is intrinsically parity-violating.
A concrete example includes, but is not limited to, the dark current induced from the axion and dark photon Chern-Simons coupling, i.e. $ j_{X}^{\mu} \propto {g_{aX}}(\partial_{\nu} a) \tilde{\hat{F}}_{X}^{\mu\nu}$ with some coupling constant $g_{aX}$.\footnote{
Having two photons, we can have three axion couplings,
\begin{align*}
 {\cal L}_{\rm int} \supset \frac{a}{f_a} \left[
 c_1 \hat{F}_{\mu\nu} \tilde{\hat{F}}^{\mu\nu} + c_2 \hat{F}_{X\mu\nu}\tilde{\hat{F}}_X^{\mu\nu}
 + c_3 \hat{F}_{\mu\nu}\tilde{\hat{F}}_X^{\mu\nu} \right] \, , 
\end{align*}
where $c_1,c_2$ and $c_3$ are, in principle, independent parameters~\cite{Lee:2015zqz}.
Here, we set $c_2\sim 1$ as the only non-zero parameter, then $c_3\sim \epsilon c_2\sim \epsilon$ and $c_1\sim \epsilon^2 c_2\sim \epsilon^2$ are induced by kinetic mixings after the diagonalization.
We note that $c_1$ is responsible for the isotropic birefringence of photon, and $c_3$ affects the anisotropic birefringence and other observables such as intensity, and circular polarization. \sml{More detailed study for this axion example is given in Appendix.~\ref{app:axion}.}
\label{footnote:axion}
}
%

\section{Polarization tensor with birefringent dark photon}
\label{section:polarization}

\begin{figure} 
\centering
\includegraphics[width=0.7\textwidth]{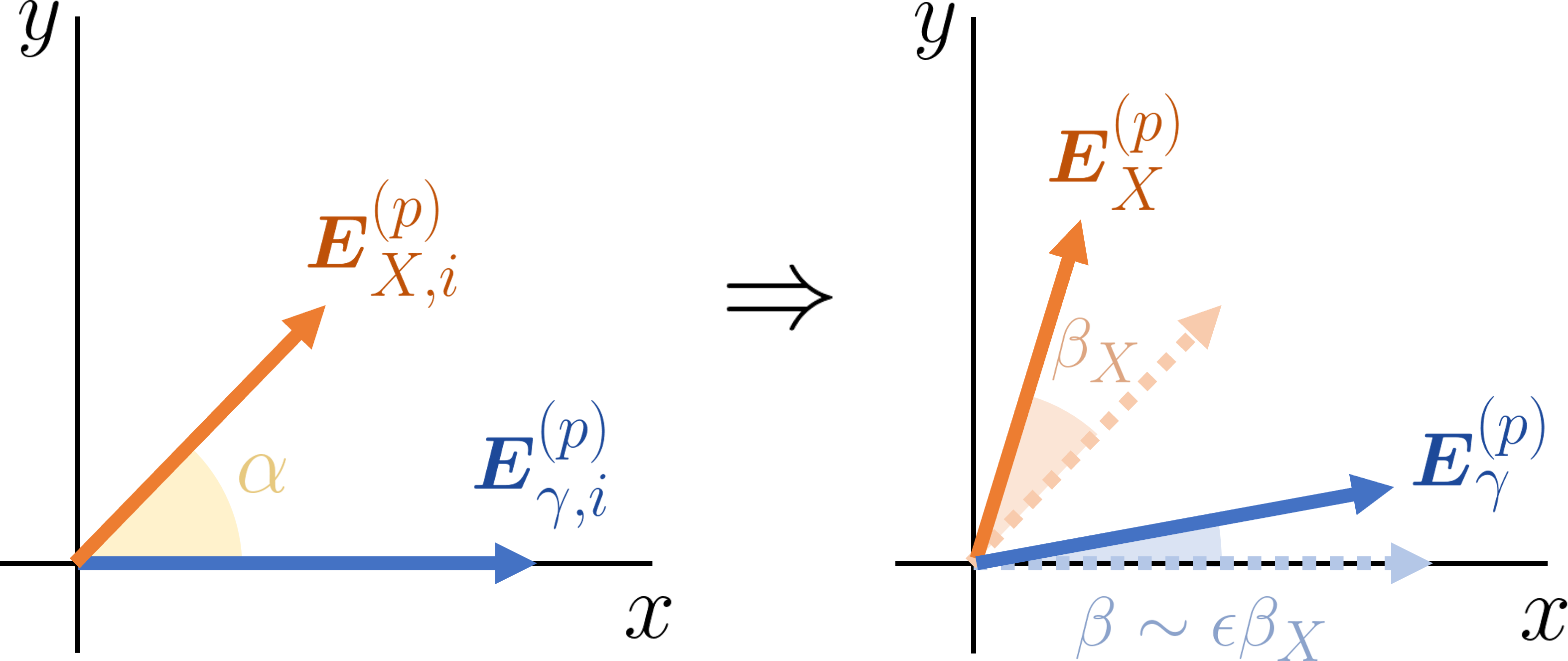}
\caption{
A schematic sketch defining the variables in the initial conditions (left) and the effects of birefringent dark photon on photon's linear polarization (right). 
\label{fig:schematic}}
\end{figure}

In the expanding universe described by a flat FLRW metric $d s^2=a(\eta)^2 (-d \eta^2 + \delta_{ij} dx^idx^j)$
with a conformal time $d\eta=dt/a(t)$, Eq.~\eqref{eq:new maxwell} implies that 
a linear combination, $ \tilde{A}^{\mu} \equiv A^{\mu} - \epsilon A_X^{\mu}$, propagates freely as a monochromatic wave in the SM vacuum $j^\nu=0$.
The photon component $A^{\mu}$ is determined by the monochromatic wave $\tilde{A}^\mu$ and {birefringent wave} $A_X^\mu$.
Therefore, the parity-violating effect appears in the visible component if the dark component is parity-violating, even though the $\epsilon$ factor suppresses the effect.
Without loss of generality, we set the direction of the propagation in the $z$-direction, and the initial amplitude of (partially) linearly polarized photon and dark photon in the $xy$-plane as, respectively,
\begin{align}
    \bm{E}_{\gamma, i}^{(p)} = \sqrt{I_{0} P} \begin{pmatrix}
        1 \\ 0
    \end{pmatrix}, &&
    \bm{E}^{(p)}_{X, i} = \sqrt{I_{X}P_{X}} e^{i\delta_{X}}\begin{pmatrix}
        \cos \alpha \\
        \sin \alpha
    \end{pmatrix}\,,  \label{eq:initial_conditions}
\end{align}
where we allow a phase factor $\delta_X$ and an angle $\alpha$ with respect to the photon.
Here, $P ~(P_X)\in [0,1]$ is the degree of polarization of photon (dark photon), and $I_{0}$ $(I_X)$ is the initial intensity of photon (dark photon).
In general, they depend on the {direction of the} line of sight $\hat{\bm{n}}$.
For a brief review of the polarization theory and related definitions, see Appendix~\ref{app:polarization}.
When the dark photon propagates through a birefringent medium (a time-varying axion medium, for instance), the polarization vector evolves into an emergent state: 
\begin{align}
\bm{E}_{X,i}^{(p)} \to  \bm{E}_{X}^{(p)} &= \hat{U}(\beta_X) \bm{E}_{X, i}^{(p)}
=\sqrt{I_{X}P_{X}} e^{i \delta_{X}}\begin{pmatrix}
  \cos(\alpha + \beta_{X}) \\ \sin(\alpha + \beta_{X})
\end{pmatrix},
\end{align}
with the rotation matrix 
$\hat{U}(\beta_X)=\begin{pmatrix}
\cos\beta_X & -\sin\beta_X  \\
\sin\beta_X &  \cos\beta_X \\
\end{pmatrix}$ induced by dark birefringence.
A concrete example is the axion coupling to dark photon, that generates $\beta_{X}\propto g_{aX} \int_{\eta_{i}}^{\eta} d\eta^{\prime} \frac{da}{d\eta}(\eta^{\prime})$.
Schematic pictures of our setup and the birefringences of photon and dark photon are depicted in Figure~\ref{fig:schematic}.
%


%
The polarization tensor is given for the partially polarized dark photon:
\begin{align}
    \rho_{X}= 
    \frac{1}{2} \begin{pmatrix}
    1 + P_{X} \cos (2\alpha + 2\beta_X)  & P_{X} \sin   (2\alpha + 2\beta_X)  \\
        P_{X} \sin  (2\alpha + 2\beta_X) &  1 - P_{X} \cos (2\alpha + 2\beta_X)
\end{pmatrix}
\end{align}
with the Stokes parameters being
$Q_{X} / I_{X} =\rho_{X11}-\rho_{X22}$, $U_{X} / I_{X} =\rho_{X12}+\rho_{X21}$ and $ V_{X} / I_{X} = i(\rho_{X12}-\rho_{X21})=0$.
We note that no circular polarization is generated from the birefringence. 
Because $\tilde{\bm{E}} = \bm{E} - \epsilon \bm{E}_X$ freely propagates, the Jones matrix constructed with this degree is time-independent, i.e. $\tilde{J}_{\alpha\beta}(t) \equiv \langle \tilde{E}_\alpha \tilde{E}^{*}_\beta \rangle_T =\tilde{J}_{\alpha\beta}(0)$.
Here, $\langle \, \cdots \, \rangle_{T}$ means taking an average over a time interval $T \gg \omega^{-1} $ where $\omega $ is the frequency of the oscillation.
As 
\begin{align}
\langle \tilde{E}\tilde{E}\rangle_T = \langle (E-\epsilon E_X)(E-\epsilon E_X)\rangle_T = \langle EE\rangle_T-\epsilon \langle EE_X +E_XE\rangle_T +\epsilon^2 \langle E_XE_X\rangle_T,    
\end{align}
the photon polarization $\langle EE\rangle_T = \langle \tilde{E}\tilde{E}\rangle_T+ \epsilon \langle EE_X+E_XE\rangle_T +{\cal O}(\epsilon^2)$ evolves with $\epsilon$.
Explicitly, by subtracting the values at $t>0$ and $t=t_{\rm ini}=0$, we have
\begin{equation}
     \begin{aligned}
   J(t) & =  J(0) +  2 \epsilon \sqrt{I_0 I_{X}} \sqrt{P P_{X}} \sin \left( \dfrac{\beta_{X}}{2} \right)  
   \\
    & \hspace{2.7cm} \times \begin{pmatrix}
        - 2 \cos \delta_{X} \sin \left( \alpha + \dfrac{\beta_{X}}{2} \right)  &  e^{-i \delta_{X}} \cos \left( \alpha + \dfrac{\beta_{X}}{2} \right)  \\
         e^{i \delta_{X}} \cos \left( \alpha + \dfrac{\beta_{X}}{2} \right)   & 0 
    \end{pmatrix} + \mathcal{O}(\epsilon^{2})
    \, , 
\end{aligned}
\end{equation}
where $J_{\alpha\beta}(t)=\langle E_\alpha E_\beta^*\rangle_{T}(t)$, and the initial tensor, given by the initial condition of the photon, is 
\begin{align}
    J_{\alpha\beta}(0)
=\dfrac{1}{2} I_0 \begin{pmatrix}
      1+P & 0 \\
        0 & 1-P
        \end{pmatrix} \equiv I_{0} \rho_{0} \, .
\end{align}   
As a direct consequence, we find that photon intensity changes inducing spectral distortion of ${\cal O}(\epsilon)$, the corresponding polarization tensor $\rho = J/I$ and the Stokes parameters for the photon are given, respectively, as
\begin{equation}
        \begin{aligned}
        \Delta I&=I-I_0= {\rm Tr}( J(t)-J(0) )   
        \\
        &= - 4 \epsilon \sqrt{I_0 I_X P P_X}
         \cos \delta_{X} 
        \sin \left( \alpha + \frac{\beta_{X}}{2} \right) \sin \left( \frac{\beta_{X}}{2} \right) + \mathcal{O}(\epsilon^{2}) \, ,
   \label{eq:intensity}
    \end{aligned}
\end{equation}
    \begin{equation}
    \begin{aligned}
    \rho
    & = \rho_{0} 
    - 2 \epsilon \sqrt{\frac{I_{X}}{I_{0}}} \sqrt{P P_{X}} \sin \left(\dfrac{\beta_{X}}{2}\right)  \\
    & \hspace{1.5cm} \times
    \begin{pmatrix}
        (1-P) \cos \delta_{X} \sin \left( \alpha + \dfrac{\beta_{X}}{2} \right)  & e^{-i \delta_{X}} \cos \left( \alpha + \dfrac{\beta_{X}}{2} \right)  
       \\
         e^{i \delta_{X}} \cos \left( \alpha + \dfrac{\beta_{X}}{2} \right)    & - (1-P) \cos \delta_{X} \sin \left( \alpha + \dfrac{\beta_{X}}{2} \right) 
    \end{pmatrix} +{\cal O}(\epsilon^2) \, ,
    \label{eq:polarization}
    \end{aligned}
    \end{equation}
and
\begin{equation}
\begin{split}
    \frac{Q}{I} & = \rho_{11} - \rho_{22} = P - 4 \epsilon (1-P) \sqrt{\frac{I_{X}}{I_{0}}} \sqrt{P P_{X}} \cos \delta_{X} \sin \left( \alpha + \frac{\beta_{X}}{2} \right) \sin \left( \frac{\beta_{X}}{2} \right)  + \mathcal{O}(\epsilon^{2})\,,  \\
    \frac{U}{I} & = \rho_{12} + \rho_{21} = 4 \epsilon  \sqrt{\frac{I_{X}}{I_{0}}} \sqrt{P P_{X}} \cos \delta_{X} \cos \left( \alpha + \frac{\beta_{X}}{2} \right) \sin \left( \frac{\beta_{X}}{2} \right)  + \mathcal{O}(\epsilon^{2})\,, \\
    \frac{V}{I} & = i (\rho_{12} - \rho_{21} ) = 4 \epsilon  \sqrt{\frac{I_{X}}{I_{0}}} \sqrt{P P_{X}} \sin \delta_{X} \cos \left( \alpha + \frac{\beta_{X}}{2} \right) \sin \left( \frac{\beta_{X}}{2} \right) + \mathcal{O}(\epsilon^{2})\, . 
\end{split}    
\label{eq:stokes}
\end{equation}

Change of intensity and polarization tensor are the key results of this article. 
Our results explicitly suggest that a birefringence effect in the photon could be induced by the polarization of the dark photon. 
%

\section{Observational Implications}
\label{section:implication}

%
In this section, we will discuss the possible observational implications. 
For a definite case study, let us assume that the phase space distribution of dark photons follows the thermal (Planck) equilibrium at the time of CMB decoupling time with a dark temperature $ T_{X} \equiv r T_{\gamma}$, which is different from the CMB temperature $T_\gamma$ in general by a factor $r$. To satisfy the $N_{\rm eff}$ constraint from Planck \cite{Planck:2018vyg}, $r$ must be smaller than 0.4 \cite{Gurian:2021qhk}.
This may be the case when the dark photons were in thermal equilibrium with dark matter at an earlier time, before the dark recombination and dark decoupling \cite{Kaplan:2009de,Fan:2013yva,Gurian:2021qhk}.
After the decoupling, the dark photon would freely stream while keeping in the phase space the Planck distribution function with reduced temperature.
Note that, to avoid the constraints coming from the lack of dark acoustic oscillation \cite{Cyr-Racine:2013fsa}, the dark decoupling should happen well ahead of the cosmic recombination at $z\simeq 1100$ \cite{Gurian:2021qhk}.

\subsection{Spectral distortion}

The spectral distortion is given in Eq.~\eqref{eq:intensity}.
While the spatial average of the random distortion vanishes, the statistical dispersion does not, even though its size is bounded from above by the kinetic mixing angle $\epsilon$ as long as $I_X < I_0$:
\begin{align}
	\frac{\delta I}{I_{0}} \simeq 2 \epsilon \sqrt{\frac{I_{X}}{I_0}} \sqrt{\bar{P} \bar{P}_{X}} \left\vert \sin  \left( \frac{\beta_{X}}{2} \right) \right\vert \lesssim 2\epsilon \, ,
 \label{Eq:SpectralDistortion}
\end{align}
where \sml{$ \delta I \equiv \sqrt{ \langle \Delta I^{2} \rangle } $ and} $\bar{P}_{(X)} \equiv \sqrt{\left\langle P_{(X)}^{2} \right\rangle}$ with $\langle \, \cdots \, \rangle $ taking an ensemble average over the sky. 
Since 
$ I
\propto {k^{3}}/\big[{e^{k/(2 \pi T)}-1}\big]$ 
for blackbody photons, we find 
\begin{align}
	\frac{I_{X}}{I_{0}}
	=
	\begin{dcases}
		r & (k \ll T_{\rm X}) \\
		\exp\left( - \frac{1-r}{r} \frac{k}{2 \pi T_{\rm \gamma}} \right) & (k \gg T_{\rm X})
	\end{dcases}\,.
	\label{eq:kdep}
\end{align}
This characteristic frequency dependence can be a smoking gun signal of the kinetic mixing, and it is distinguishable from the spectral distortion due to the axion decays when $m_a \in {\rm (keV, MeV)}$~\cite{Balazs:2022tjl}.\footnote{\sml{Even though we only consider the massless dark photon, in the case of the massive dark photon for certain mass ranges, there are other possible impacts on spectral distortion \cite{Mirizzi:2009iz,Caputo:2020bdy,Caputo:2020rnx}, or CMB anisotropies \cite{Aramburo-Garcia:2024cbz}.}}
At high frequencies with $\hbar \omega\gg 3k_B T_{X}$ in the Wien tail, the intensity of dark photon is suppressed thus the effect on the CMB polarization is minuscule. 
{Current bounds on the spectral distortion is $\mathcal{O}(10^{-5})$ and expected to be improved up to $\mathcal{O}(10^{-8})$~\cite{Kogut:2019vqh}.}

\begin{figure}[t]
    \centering
    \includegraphics[width=0.7\textwidth]{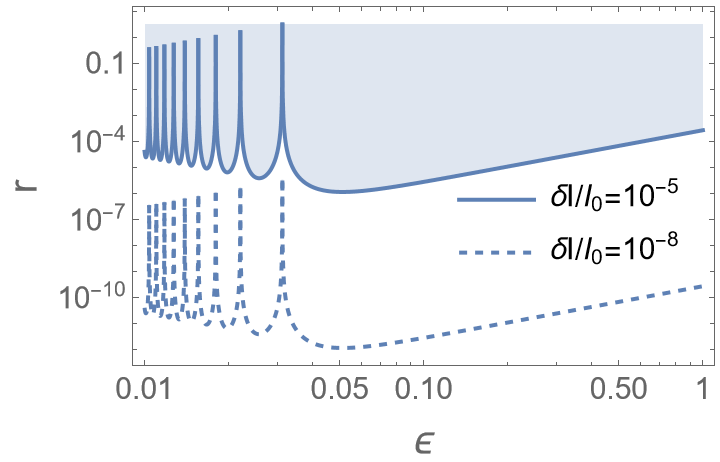}
    \caption{
    {Constraints on the temperature of the dark sector ($r \equiv T_{X}/T_{\gamma} $) from the current/future constraints/sensitivity on the spectral distortion for simultaneous explanation of the isotropic birefringence at $\epsilon^{2}$ order as $ \beta_{\rm iso} = \epsilon^{2} \beta_{X} $. We set $ \sqrt{\bar{P} \bar{P}_{X}} = 0.1 $ for a definite illustration.}}
    \label{fig:intensity}
\end{figure}

From the Eq.~\eqref{Eq:SpectralDistortion}, we can see nontrivial implications on the intensity and degree of polarization within the dark sector: For instance, assuming that there exists isotropic birefringence $\beta_{\rm iso} \simeq \epsilon^{2} \beta_{X} $ and this explains the recent observation \cite{Minami:2020odp,Diego-Palazuelos:2022dsq,Eskilt:2022wav,Eskilt:2022cff}, the same parameter space should have suppressed value of $I_{X}$ or $\bar{P}_{X}$. As a definite illustration, the bound on $r$ with $ \sqrt{\bar{P} \bar{P}_{X}} = 0.1$ is depicted in Figure~\ref{fig:intensity} using the current and future constraints on the spectral distortion: $\delta I/I_0=10^{-5}$ and $10^{-8}$, respectively.

\subsection{Birefringence}

Non-zero value of $U$ in Eq.~\eqref{eq:stokes} implies that there exists birefringence in CMB as 
\begin{align}
          &\beta (\hat{\bm{n}}) = \frac{1}{2} \arctan \left(\frac{U}{Q}\right)  
    = 2 \epsilon \sqrt{ \frac{I_{X} P_{X}}{I_0 P}} \cos \delta_{X} \cos \left( \alpha + \frac{\beta_{X}}{2} \right) \sin \left( \frac{\beta_{X}}{2} \right) +{\cal O}(\epsilon^2)\, . \label{eq:birefringence}
\end{align}
Note that, in general, the angle $\alpha$ is unknown. For example, if dark recombination \cite{Gurian:2021qhk} happened, as for the atomic dark matter model \cite{Kaplan:2009de}, the linear polarization of dark photons is determined by the local quadrupole at the dark recombination time, which must be earlier than the cosmic recombination time \cite{Cyr-Racine:2013fsa}.
{Taking the random $\alpha$ over the sky, 
we expect the monopole, a constant isotropic birefringence angle, to appear as
\begin{align}
\beta_{\rm iso} \equiv \langle \beta( \hat{\bm{n}} ) \rangle \simeq \epsilon^2 \beta_{X},
\label{eq:beta_iso}
\end{align}
due to the effective photon-axion coupling induced by the photon-dark photon mixings. 
We note that the recently reported cosmic birefringence at 3.6-$\sigma$ level of $\beta_{\rm iso} \sim 0.35^{\circ} \simeq 6.1\times 10^{-3} $~\cite{Minami:2020odp,Diego-Palazuelos:2022dsq,Eskilt:2022wav,Eskilt:2022cff} could be accounted with $\epsilon \approx 0.078$ and $\beta_{X} = 1$, for example.}

On the other hand, the variance appears as 
\begin{align}
\left\langle \beta^{2}_{\rm aniso} \right\rangle 
&  \simeq \epsilon^2 \frac{I_{X}}{I_{0}} \left\langle \frac{P_{X}}{P} \right\rangle  \sin^{2} \left( \frac{\beta_{X}}{2} \right).
\end{align}
The variance can affect the $TE$-correlation as ~\cite{Li:2008tma,Gluscevic_2012}
\begin{align}
    C_{\ell}^{TE} 
    \rightarrow
    C_{\ell}^{TE} \cos (2\beta_{\rm iso}) 
    \left( 1- 2 \langle \beta_{\rm aniso}^{2} \rangle \right) \, .
\end{align}
Similarly, $C^{EE}_\ell$- and $C^{BB}_\ell$-correlations, and higher order correlations  $C^{EBEB}_\ell$ are all affected by $\beta_{\rm aniso}$. 
The currently available bound for the variance of birefringence is $ \langle \beta_{\rm aniso}^{2} \rangle \lesssim  10^{-5}$~\sml{\cite{Mei:2014iaa,Pan:2016vai,Contreras:2017sgi,Gruppuso:2020kfy,SPT:2020cxx,Namikawa:2020ffr,Bortolami:2022whx,Zagatti:2024jxm}}.

\subsection{Circular polarization}

The non-vanishing circular polarization, called the CMB $V$-mode, is predicted in Eq.~\eqref{eq:stokes}:
\begin{align}
\langle V^{2} \rangle \simeq 4 \epsilon^{2} I_{0} I_{X}  \bar{P} \bar{P}_{X} \sin^{2} \left( \frac{\beta_{X}}{2} \right).
\end{align}
We emphasize that the non-vanishing $V\sim U$ is a characteristic feature of our model with kinetic mixing. {Only a negligible amount of circular polarization is generated from the axion decay~\cite{Finelli:2008jv, Alexander:2008fp}.}
The currently allowed circular polarization is \sml{up to $\mathcal{O}(1 \mu \text{K})$~\cite{SPIDER:2017kwu,Padilla:2019dhz}}.

{We summarize the difference between the dark photon model with a coupling to a pseudoscalar $g_{\theta X} \theta F_{X \mu\nu} \tilde{F}_{X}^{\mu\nu}  $ and mixing to the photon, and the axion model which the pseudo scalar directly couples to the photon with $g_{\theta} \theta F_{ \mu\nu} \tilde{F}^{\mu\nu}  $ in Table~\ref{table:comparison}. We assume there exists field excursion $\Delta \theta$, and this should be homogeneous for isotropic birefringence.}

\begin{table}
\begin{center}
\begin{threeparttable}
\begin{tabular}{l | c c }
\hline & \textbf{Dark Photon} & \textbf{Axion} \\
\hline
 Isotropic Birefringence &  ${\cal O}(\epsilon^2 g_{\theta X} \Delta \theta)$ &  ${\cal O}(g_\theta \Delta \theta)\tnote{$*$} $\\
 \hline
 Anisotropic Birefringence & ${\cal O}(\epsilon g_{\theta X} \Delta \theta )$ &  ${\cal O}(g_\theta \Delta \theta)$\\
 \hline
  Spectral Distortion &  Yes  &  Yes$\tnote{$**$} $  \\
  \hline 
Circular Polarization & ${\cal O}(\epsilon g_{ \theta X } \Delta \theta)$ & Negligible \\
\hline
\end{tabular}
\begin{tablenotes}\footnotesize
\item[$*$] For a mass window smaller than the Hubble scale of today $H_{0} \sim 10^{-33}~\text{eV}$.
\item[$**$] Only in a mass window (keV-MeV)~\cite{Balazs:2022tjl}.
\end{tablenotes}
\end{threeparttable}
    \caption{Schematic comparison of observational signatures from birefringent dark photon mixing considered in this work and direct coupling to pseudoscalar $\theta$ with field value difference $\Delta \theta$. Note that the prefactors, influenced by the intensities and polarization degrees of both the photon and dark photon, are not explicitly written in this table. For a more detailed explanation, consult the main text.
\label{table:comparison}}
\end{center}
\end{table}

\section{Conclusions}
\label{section:conclusion}

Recent CMB observations have hinted at the cosmic birefringence in tantalizing 3.6-$\sigma$ level. In literature, the birefringence is usually attributed to the direct coupling between photons and a pseudoscalar field, like an axion. Here, we investigate an alternative explanation with the dark photon.

Our main results are encapsulated in Eq.~\eqref{eq:intensity} and Eq. \eqref{eq:polarization}, which represent, respectively, the intensity and polarization tensor of the photon kinetically mixed to the birefringent dark photon.
We find that the kinetic mixing not only transfers the dark photon's birefringence to the CMB photon, but also yields the unique, distinctive features such as direction-dependent spectral distortion and circular polarization.
The birefringence from the kinetic mixing is anisotropic birefringence with non-zero variance.

Finally, we acknowledge that exploring a more concrete model of parity violation in the dark sector would open up broader theoretical possibilities and potentially uncover new avenues for observations of BSM physics.
We leave these tasks to future work.

\vspace{1.0cm}

\acknowledgments

We thank Gongjun Choi, Johannes Eskilt, Kunio Kaneta, Suro Kim, Kazunori Kohri and Marko Simonovic for helpful discussions. 
This work is supported in part by the National Research Foundation grants 2019R1A2C2085023 (JG), 2021R1A2B5B02087078, 2021R1A2C2011003 and 
RS-2023-00246268 (DWJ), RS-2023-00283129 and RS-2024-00340153 and by Yonsei internal grant for Mega-science (2023-22-048). (SCP).
SML was also supported in part by the Hyundai Motor Chung Mong-Koo Foundation Scholarship, and the Korea-CERN Theoretical Physics Collaboration and Developing Young High-Energy Theorists Fellowship Program (NRF-2012K1A3A2A0105178151).
DWK is supported in part by KIAS Individual Grant PG076202.
JG is further supported in part by the Korea-Japan Basic Scientific Cooperation Program supported by the National Research Foundation of Korea and the Japan Society for the Promotion of Science (NRF-2020K2A9A2A08000097), and
the Ewha Womans University Research Grant of 2022 (1-2022-0606-001-1) and 2023 (1-2023-0748-001-1).
DJ is supported by KIAS Individual Grant PG088301 at Korea Institute for Advanced Study.
DWJ was also supported in part by IBS under the project code IBS-R018-D3.
SCP, SML and DWK are grateful to the CERN for hospitality while this work was under progress.
SCP and JG are also grateful to the Asia Pacific Center for Theoretical Physics for their hospitality while this work was in progress.
JG and DJ thank the Yukawa Institute for Theoretical Physics at Kyoto University, where this work was completed during the YITP-T-23-03 ``Revisiting cosmological non-linearities in the era of precision surveys.''


\appendix

\section{Review of (Partially) Polarized Light}
\label{app:polarization}

In this appendix, we briefly review the theory of partially polarized light. Especially, we will explicitly show how the polarization tensor is defined and set the notations. This part mainly relies on Ref.~\cite{Landau:1975pou}.
A electric field at a fixed position $\bm{x} = 0$ is given as 
$ \bm{E} (t) e^{- i \omega t} $ for a fixed $\bm{k}$ with $\vert \bm{k} \vert = \omega$.
Here, $\bm{E}$ can have a time dependence in general and is decomposed into polarized part $(p)$ and unpolarized (or natural) part $(n)$ as
\begin{align}
    \bm{E} = \bm{E}^{(p)} + \bm{E}^{(n)} \, . 
\end{align}
The unpolarized part satisfies
\begin{align}
    \left\langle E_{\alpha}^{(n)} E_{\beta}^{(n) *} \right\rangle_{T} = \frac{1}{2} I^{(n)} \delta_{\alpha\beta}
    \, ,
\end{align}
where $\langle \, \cdots \, \rangle_{T}$ is an average over a time interval $T$ much larger than $\omega^{-1}$.
Here, $I^{(n)}$ is the intensity of the unpolarized part of the electric field. On the other hand, the polarized part is assumed to be nearly constant compared to the time scale of the average.
With this decomposition, we define the Jones matrix
\begin{align}
    J_{\alpha\beta} \equiv \left\langle E_{\alpha} E_{\beta}^{*} \right\rangle_{T}
    = E_{\alpha}^{(p)} E_{\beta}^{(p)*}  + \frac{1}{2} I^{(n)} \delta_{\alpha\beta}
    \,,
\end{align}
the intensity $ I \equiv \text{Tr} \, J $ and polarization tensor $\rho_{\alpha\beta} \equiv J_{\alpha\beta} / I $.
If a quantity is slowly varying in a much larger time scale than the time scale of averaging, there may be residual time dependence after averaging fast modes.
This also includes the observational effects we discuss in the main text arouse from the slow birefringence of the dark photon $\beta_{X}(t)$.
In terms of Stokes parameters, the polarization tensor is written as
\begin{align}
    \rho = \frac{1}{2 I} \begin{pmatrix}
        I + Q & U - i V \\
      U + i V  & I - Q
    \end{pmatrix}
    \, .
\end{align}
We also introduce the degree of the polarization $P \in [0,1]$ as $\det \rho \equiv (1-P^{2})/4$ where $P = 0$ corresponds to the unpolarized light, and $P=1$ is for the completely polarized one.
With these definitions, $I^{(n)} = I (1-P)$, and $I^{(p)} \equiv \vert \bm{E}^{(p)} \vert^{2} = I P $. Also, $P = \sqrt{Q^{2} + U^{2} + V^{2}} / I$.

\section{Birefringence from Axion with Two Photons}
\label{app:axion}

In this appendix, we present a more detailed examination of the axion example. While the main text remains agnostic about the source of birefringence in the dark photon, ensuring that our results are broadly applicable, this appendix focuses specifically on the axion. By doing so, we can highlight some unique features that emerge in the axion scenario.

We begin with the general Lagrangian for two photons allowing finite kinetic mixing and the general axionic couplings ($\hat{c}_{i=1,2,3}$): 
\begin{equation}
    \begin{aligned}
    \mathcal{L} & = - \frac{1}{4} \hat{F}_{\mu\nu}  \hat{F}^{\mu\nu} -\frac{1}{4} \hat{F}_{X\mu\nu}  \hat{F}_{X}^{\mu\nu} -\frac{\varepsilon}{2} \hat{F}_{\mu\nu}  \hat{F}_{X}^{\mu\nu} + \frac{\theta}{4} \left[ \hat{c}_{1} \hat{F}_{\mu\nu} \tilde{\hat{F}}^{\mu\nu} +  \hat{c}_{2} \hat{F}_{X\mu\nu} \tilde{\hat{F}}_{X}^{\mu\nu} 
    + 2 \hat{c}_{3} \hat{F}_{\mu\nu} \tilde{\hat{F}}_{X}^{\mu\nu}   
    \right] \\
    & \rightarrow  - \frac{1}{4} F_{\mu\nu}  F^{\mu\nu} -\frac{1}{4} F_{X\mu\nu}  F_{X}^{\mu\nu} + \frac{\theta}{4} \left[ c_{1} F_{\mu\nu} \tilde{F}^{\mu\nu} +  c_{2} F_{X\mu\nu} \tilde{F}_{X}^{\mu\nu} 
    + 2 c_{3} F_{\mu\nu} \tilde{F}_{X}^{\mu\nu}   
    \right]
\end{aligned}
\end{equation}
where 
\begin{equation}
    \begin{aligned}
    c_{1} & = \frac{1}{1 - \varepsilon^{2}} \left(  \hat{c}_{1} - 2 \varepsilon \hat{c}_{3} + \varepsilon^{2} \hat{c}_{2}  \right)  = \hat{c}_{1} - 2 \varepsilon \hat{c}_{3} + \varepsilon^{2} (\hat{c}_{1} + \hat{c}_{2})  + \mathcal{O}(\varepsilon^{3}) \,, \\   
    c_{2} & = \hat{c}_{2}  \,,  \\
    c_{3} & = \frac{1}{\sqrt{1-\varepsilon^{2}}} \left( \hat{c}_{3} - \varepsilon \hat{c}_{2} \right) = \hat{c}_{3} - \varepsilon \hat{c}_{2} + \frac{\varepsilon^{2}}{2} \hat{c}_{3} + \mathcal{O}(\varepsilon^{3}) \,,
    \end{aligned}
\end{equation}
after the field redefinition in Eq.~\eqref{eq:rotation} where the approximation holds for small mixing.  



The case we consider in the main text corresponds to $(\hat{c}_1,\hat{c}_2,\hat{c}_3)=(0,1,0)$ and $\varepsilon\ll 1$. Hereafter, we will use $ \epsilon \equiv - \varepsilon \ll 1 $ after setting $ e = e_{X} = 1$ for brevity.

\subsection{General Solution}

With non-vanishing $c_1,c_2$ and $c_3$, we have generalized equations for two photons:
\begin{equation} 
    \begin{aligned}
    \partial_{\nu} F^{\mu\nu} & = (\partial_{\nu} \theta) \left[ c_{1} \tilde{F}^{\mu\nu} + c_{3} \tilde{F}_{X}^{\mu\nu} \right], \\
    \partial_{\nu} F_{X}^{\mu\nu} & = (\partial_{\nu} \theta) \left[ c_{2} \tilde{F}_{X}^{\mu\nu} + c_{3} \tilde{F}^{\mu\nu} \right].
    \end{aligned}
\end{equation}
It is convenient to write the equations in a matrix notation,
\begin{align}
    \partial_{\nu} \vec{F}^{\mu\nu} =  (\partial_{\nu} \theta) C \tilde{ \vec{ F } }^{\mu\nu}, && \vec{F} = \begin{pmatrix}
        F \\
        F_{X}
    \end{pmatrix}, && C = \begin{pmatrix}
        c_{1} & c_{3} \\
        c_{3} & c_{2}
    \end{pmatrix} \, .
\end{align}
In general, $ C $ is diagonalizable unless $ \det C \neq 0 $,
and we have
\begin{align}
    C = X D X^{-1}
\end{align}
where $ D = \text{diag} (\lambda_{1}, \lambda_{2} )$ with
\begin{equation}
    \begin{aligned}
    \lambda_{1} & = \frac{1}{2} \left( c_{1} + c_{2} - \sqrt{(c_{1} - c_{2})^{2} + 4 c_{3}^{2}} \right) \, , \\
    \lambda_{2} & =  \frac{1}{2} \left( c_{1} + c_{2} + \sqrt{(c_{1} - c_{2})^{2} + 4 c_{3}^{2}} \right) 
    \end{aligned}
\end{equation}
and $ X = ( \vec{x}_{1} \vert \vec{x}_{2} ) $ with $ \vec{x}_{1,2} $ being corresponding orthonormal eigenvectors.

Then,
\begin{align}
    \partial_{\nu} (X^{-1} \vec{F})^{\mu\nu} = (\partial_{\nu} \theta) D (X^{-1} \tilde{ \vec{F} }^{\mu\nu}  )
\end{align}
or
\begin{align}
\partial_\nu \mathcal{F}_{\alpha}^{\mu\nu}= (\partial_\nu\theta) \lambda_{\alpha} \tilde{ \mathcal{F} }_{\alpha}^{\mu\nu},
\end{align}
where $ \mathcal{F}_{\alpha} :=(X^{-1}\vec{ F })_{\alpha} $ ($\alpha = 1, 2$). Therefore, we have two independently birefringent fields $ \mathcal{F}_1 $ and $ \mathcal{F}_2 $ with birefringence angles accumulated by the line of sight (LOS) integral:
\begin{align}
\beta_{\alpha} = \frac{ \lambda_{\alpha} }{2}  
\int_{\rm LOS} d\theta \equiv \frac{ \lambda_{\alpha} }{2}  \Delta \theta.
\end{align}
Finally, the original fields, $F_{\alpha=1,2}$ ($ F_{1} = F $ and $F_{2} = F_{X}$), are recovered by $F_\alpha=(X \vec{\mathcal{F}} )_{\alpha}$ with $ \vec{\mathcal{F}} = (\mathcal{F}_1, \mathcal{F}_{2})^{T}$.

\subsection{Case Study (1) : $ c_{1,2} \sim 1 \gg c_{3} \sim \epsilon $}

Without loss of generality, we set $ c_{2} > c_{1} $ and $ c_{3} \equiv \epsilon \tilde{c}_{3} > 0 $. The $X$ matrix is explicitly given 
\begin{align}
    X =  
    \begin{pmatrix}
         1 - \dfrac{\epsilon^{2} \tilde{c}_{3}}{2 (c_{1} - c_{2})^{2}}  &  - \dfrac{\epsilon \tilde{c}_{3}}{c_{1} - c_{2}} \\
         \dfrac{\epsilon \tilde{c}_{3}}{c_{1} - c_{2}} &  1 - \dfrac{\epsilon^{2} \tilde{c}_{3}}{2 (c_{1} - c_{2})^{2}}
    \end{pmatrix} + \mathcal{O}(\epsilon^{3}) \, ,  
\end{align}
and the mixing matrix is diagonalized as 
\begin{align}
    D = \begin{pmatrix}
         c_{1} + \dfrac{ \epsilon^{2} \tilde{c}_{3}^{2}}{c_{1} - c_{2}}  & \\
        & c_{2} - \dfrac{ \epsilon^{2} \tilde{c}_{3}^{2}}{c_{1} - c_{2}}  
    \end{pmatrix} + \mathcal{O}(\epsilon^{3})  \equiv \begin{pmatrix}  \lambda_{1} & \\ 
    & \lambda_{2} 
    \end{pmatrix} \, .
\end{align}
The field $ \mathcal{F}_{\alpha} $, in terms of the electric components, evolves as
\begin{align}
     X^{-1} \begin{pmatrix}
         \bm{E}_{\gamma,i}^{(p)} \\ \bm{E}_{X,i}^{(p)}
     \end{pmatrix}  \rightarrow 
    \begin{pmatrix}
        \hat{U}(\beta_{1}) \left[ (X^{-1})_{11} \bm{E}_{\gamma,i}^{(p)} + (X^{-1})_{12} \bm{E}_{X,i}^{(p)} \right] \\
        \hat{U}(\beta_{2}) \left[ (X^{-1})_{21}^{(p)} \bm{E}_{\gamma,i}^{(p)} + (X^{-1})_{22}^{(p)} \bm{E}_{X,i}^{(p)} \right].
    \end{pmatrix}
\end{align}
Now, the solution for the photon is obtained with the initial conditions given in Eq.~\eqref{eq:initial_conditions} with $ E_{\gamma,0}^{(p)} = \sqrt{I_{0}P}$ and $ E_{X,0}^{(p)} = \sqrt{I_{X}P_{X}}$:  
\begin{equation}
    \begin{aligned}
    \bm{E}_{\gamma}^{(p)} & = E_{\gamma,0}^{(p)} \begin{pmatrix}
        \cos \beta_{1} \\
        \sin \beta_{1} 
    \end{pmatrix}  + \epsilon  \frac{\tilde{c}_{3}}{c_{1} - c_{2}} E_{X,0}^{(p)} e^{i \delta} \begin{pmatrix} \cos \left( \alpha +  \beta_{1} \right) - \cos (\alpha + \beta_{2} ) \\
        \sin \left( \alpha +  \beta_{1} \right) - \sin (\alpha + \beta_{2} ) \end{pmatrix} \\
     & \hspace{4cm} - \epsilon^{2} \frac{\tilde{c}_{3}^{2}}{(c_{1} - c_{2})^{2}} E_{\gamma,0}^{(p)}\begin{pmatrix} \cos  \beta_{1}  - \cos \beta_{2}  \\
        \sin \beta_{1} - \sin  \beta_{2}  \end{pmatrix} + \mathcal{O}(\epsilon^{3}). \label{eq:solution}
\end{aligned}
\end{equation}
With this constructed solution, we finally obtain the birefringence angle using the Stokes parameters:
\begin{align}
    \beta = \beta_{1} + 2  \epsilon \frac{\tilde{c}_{3}}{c_{1} - c_{2}} \frac{E_{X,0}^{(p)}}{E_{\gamma,0}^{(p)}} \cos \delta \cos \left( \alpha - \frac{\beta_{1} - \beta_{2}}{2}  \right) \sin \left( \frac{
    \beta_{1} - \beta_{2}}{2} \right) + \mathcal{O}(\epsilon^{2}).
    \label{eq:beta_general}
\end{align}
The first term represents the isotropic birefringence due to the presence of $c_{1}$, which is independent of the dark photon. The second term accounts for the anisotropic birefringence induced by photon-dark photon mixing. It is important to note that the second term exhibits a non-linear dependence not only on $\beta_2$, and $\alpha$, but also on $\beta_1$. Consequently, the total birefringence angle is not simply the sum of the two contributions.\footnote{We thank the referee for encouraging us to explicitly verify this interesting fact.}

\subsection{Case Study (2) : $ c_{1} \sim \epsilon^2$, $c_2 \sim 1$ and $c_3\sim \epsilon$}

This case is realized when we set $ \hat{c}_{1} = \hat{c}_{3} = 0 $ and $ \hat{c}_{2} = 1 $ as we did in the main text.  Expanding the result in Eq.~\eqref{eq:beta_general} keeping the results at $\epsilon^{2}$ order, we obtain
\begin{equation}
    \begin{aligned}
  \beta & = 2 \epsilon \frac{\tilde{c}_{3}}{c_{2}} \frac{E_{X,0}^{(p)}}{E_{\gamma,0}^{(p)}} \cos \delta \cos \left( \alpha + \frac{\beta_{2}}{2} \right) \sin \left( \frac{\beta_{2}}{2} \right) \\ 
  & \quad\quad + \epsilon^{2} \left(  \tilde{\beta}_{1} + \frac{\tilde{c}_{3}^{2}}{c_{2}^{2}} \sin \beta_{2} + 2 \frac{\tilde{c}_{3}^{2}}{c_{2}^{2}} \left(\frac{E_{X,0}^{(p)}}{E_{\gamma,0}^{(p)}}\right)^{2} \cos 2 \delta \sin^{2} \left( \frac{ \beta_{2}}{2} \right) \sin \left( 2\alpha + \beta_{2} \right)  \right) + \mathcal{O}(\epsilon^{3}).
    \end{aligned}
\end{equation}
The isotropic part is
\begin{align}
    \beta_{\rm iso} = \epsilon^{2} \left( \tilde{\beta}_{1} + \frac{\tilde{c}_{3}^{2}}{c_{2}^{2}} \sin \beta_{2} \right),
\end{align}
where $ \beta_{1} = \epsilon^{2} \tilde{\beta}_{1} $ is determined by $ \lambda_{1} \simeq \epsilon^{2} (\tilde{c}_{1} c_{2} - \tilde{c}_{3}^{2}) / c_{2}  $. It is noteworthy that when $ \tilde{c}_{1} = \tilde{c}_{3} = 1 $ the first term vanishes, resulting in $\beta_{\rm iso} = \epsilon^{2} \sin \beta_{X} \approx \epsilon^2 \beta_X$ when $ \beta_{2} = \beta_{X}\ll 1$ recovering the result in Eq.~\eqref{eq:beta_iso}.

\bibliographystyle{JHEP}
\bibliography{refs}

\providecommand{\href}[2]{#2}\begingroup\raggedright\begin{thebibliography}{10}

\bibitem{Carroll:1989vb}
S.M.~Carroll, G.B.~Field and R.~Jackiw, \emph{{Limits on a Lorentz and Parity
  Violating Modification of Electrodynamics}},
  \href{https://doi.org/10.1103/PhysRevD.41.1231}{\emph{Phys. Rev. D}
  {\bfseries 41} (1990) 1231}.

\bibitem{Carroll:1991zs}
S.M.~Carroll and G.B.~Field, \emph{{The Einstein equivalence principle and the
  polarization of radio galaxies}},
  \href{https://doi.org/10.1103/PhysRevD.43.3789}{\emph{Phys. Rev. D}
  {\bfseries 43} (1991) 3789}.

\bibitem{Harari:1992ea}
D.~Harari and P.~Sikivie, \emph{{Effects of a Nambu-Goldstone boson on the
  polarization of radio galaxies and the cosmic microwave background}},
  \href{https://doi.org/10.1016/0370-2693(92)91363-E}{\emph{Phys. Lett. B}
  {\bfseries 289} (1992) 67}.

\bibitem{Weinberg:1977ma}
S.~Weinberg, \emph{{A New Light Boson?}},
  \href{https://doi.org/10.1103/PhysRevLett.40.223}{\emph{Phys. Rev. Lett.}
  {\bfseries 40} (1978) 223}.

\bibitem{Wilczek:1977pj}
F.~Wilczek, \emph{{Problem of Strong $P$ and $T$ Invariance in the Presence of
  Instantons}}, \href{https://doi.org/10.1103/PhysRevLett.40.279}{\emph{Phys.
  Rev. Lett.} {\bfseries 40} (1978) 279}.

\bibitem{Choi:2020rgn}
K.~Choi, S.H.~Im and C.~Sub~Shin, \emph{{Recent Progress in the Physics of
  Axions and Axion-Like Particles}},
  \href{https://doi.org/10.1146/annurev-nucl-120720-031147}{\emph{Ann. Rev.
  Nucl. Part. Sci.} {\bfseries 71} (2021) 225}
  [\href{https://arxiv.org/abs/2012.05029}{{\ttfamily 2012.05029}}].

\bibitem{Rees:1968}
M.J.~Rees, \emph{{Polarization and Spectrum of the Primeval Radiation in an
  Anisotropic Universe}}, \href{https://doi.org/10.1086/180208}{\emph{Astoph.
  J. Lett.} {\bfseries 153} (1968) L1}.

\bibitem{Kosowsky:1994cy}
A.~Kosowsky, \emph{{Cosmic microwave background polarization}},
  \href{https://doi.org/10.1006/aphy.1996.0020}{\emph{Annals Phys.} {\bfseries
  246} (1996) 49} [\href{https://arxiv.org/abs/astro-ph/9501045}{{\ttfamily
  astro-ph/9501045}}].

\bibitem{Zaldarriaga:1996xe}
M.~Zaldarriaga and U.~Seljak, \emph{{An all sky analysis of polarization in the
  microwave background}},
  \href{https://doi.org/10.1103/PhysRevD.55.1830}{\emph{Phys. Rev. D}
  {\bfseries 55} (1997) 1830}
  [\href{https://arxiv.org/abs/astro-ph/9609170}{{\ttfamily
  astro-ph/9609170}}].

\bibitem{Kamionkowski:1996ks}
M.~Kamionkowski, A.~Kosowsky and A.~Stebbins, \emph{{Statistics of cosmic
  microwave background polarization}},
  \href{https://doi.org/10.1103/PhysRevD.55.7368}{\emph{Phys. Rev. D}
  {\bfseries 55} (1997) 7368}
  [\href{https://arxiv.org/abs/astro-ph/9611125}{{\ttfamily
  astro-ph/9611125}}].

\bibitem{Lue:1998mq}
A.~Lue, L.-M.~Wang and M.~Kamionkowski, \emph{{Cosmological signature of new
  parity violating interactions}},
  \href{https://doi.org/10.1103/PhysRevLett.83.1506}{\emph{Phys. Rev. Lett.}
  {\bfseries 83} (1999) 1506}
  [\href{https://arxiv.org/abs/astro-ph/9812088}{{\ttfamily
  astro-ph/9812088}}].

\bibitem{Gluscevic:2010vv}
V.~Gluscevic and M.~Kamionkowski, \emph{{Testing Parity-Violating Mechanisms
  with Cosmic Microwave Background Experiments}},
  \href{https://doi.org/10.1103/PhysRevD.81.123529}{\emph{Phys. Rev. D}
  {\bfseries 81} (2010) 123529}
  [\href{https://arxiv.org/abs/1002.1308}{{\ttfamily 1002.1308}}].

\bibitem{Minami:2020odp}
Y.~Minami and E.~Komatsu, \emph{{New Extraction of the Cosmic Birefringence
  from the Planck 2018 Polarization Data}},
  \href{https://doi.org/10.1103/PhysRevLett.125.221301}{\emph{Phys. Rev. Lett.}
  {\bfseries 125} (2020) 221301}
  [\href{https://arxiv.org/abs/2011.11254}{{\ttfamily 2011.11254}}].

\bibitem{Diego-Palazuelos:2022dsq}
P.~Diego-Palazuelos et~al., \emph{{Cosmic Birefringence from the Planck Data
  Release 4}},
  \href{https://doi.org/10.1103/PhysRevLett.128.091302}{\emph{Phys. Rev. Lett.}
  {\bfseries 128} (2022) 091302}
  [\href{https://arxiv.org/abs/2201.07682}{{\ttfamily 2201.07682}}].

\bibitem{Eskilt:2022wav}
J.R.~Eskilt, \emph{{Frequency-dependent constraints on cosmic birefringence
  from the LFI and HFI Planck Data Release 4}},
  \href{https://doi.org/10.1051/0004-6361/202243269}{\emph{Astron. Astrophys.}
  {\bfseries 662} (2022) A10}
  [\href{https://arxiv.org/abs/2201.13347}{{\ttfamily 2201.13347}}].

\bibitem{Eskilt:2022cff}
J.R.~Eskilt and E.~Komatsu, \emph{{Improved constraints on cosmic birefringence
  from the WMAP and Planck cosmic microwave background polarization data}},
  \href{https://doi.org/10.1103/PhysRevD.106.063503}{\emph{Phys. Rev. D}
  {\bfseries 106} (2022) 063503}
  [\href{https://arxiv.org/abs/2205.13962}{{\ttfamily 2205.13962}}].

\bibitem{Komatsu:2022nvu}
E.~Komatsu, \emph{{New physics from the polarized light of the cosmic microwave
  background}}, \href{https://doi.org/10.1038/s42254-022-00452-4}{\emph{Nature
  Rev. Phys.} {\bfseries 4} (2022) 452}
  [\href{https://arxiv.org/abs/2202.13919}{{\ttfamily 2202.13919}}].

\bibitem{Huh:2007zw}
J.-H.~Huh, J.E.~Kim, J.-C.~Park and S.C.~Park, \emph{{Galactic 511 keV line
  from MeV milli-charged dark matter}},
  \href{https://doi.org/10.1103/PhysRevD.77.123503}{\emph{Phys. Rev. D}
  {\bfseries 77} (2008) 123503}
  [\href{https://arxiv.org/abs/0711.3528}{{\ttfamily 0711.3528}}].

\bibitem{Park:2012xq}
J.-C.~Park and S.C.~Park, \emph{{Radiatively decaying scalar dark matter
  through U(1) mixings and the Fermi 130 GeV gamma-ray line}},
  \href{https://doi.org/10.1016/j.physletb.2012.12.035}{\emph{Phys. Lett. B}
  {\bfseries 718} (2013) 1401}
  [\href{https://arxiv.org/abs/1207.4981}{{\ttfamily 1207.4981}}].

\bibitem{Fabbrichesi:2020wbt}
M.~Fabbrichesi, E.~Gabrielli and G.~Lanfranchi, \emph{{The Dark Photon}},
  \href{https://arxiv.org/abs/2005.01515}{{\ttfamily 2005.01515}}.

\bibitem{Vogel:2013raa}
H.~Vogel and J.~Redondo, \emph{{Dark Radiation constraints on minicharged
  particles in models with a hidden photon}},
  \href{https://doi.org/10.1088/1475-7516/2014/02/029}{\emph{JCAP} {\bfseries
  02} (2014) 029} [\href{https://arxiv.org/abs/1311.2600}{{\ttfamily
  1311.2600}}].

\bibitem{Berlin:2022hmt}
A.~Berlin, J.A.~Dror, X.~Gan and J.T.~Ruderman, \emph{{Millicharged relics
  reveal massless dark photons}},
  \href{https://doi.org/10.1007/JHEP05(2023)046}{\emph{JHEP} {\bfseries 05}
  (2023) 046} [\href{https://arxiv.org/abs/2211.05139}{{\ttfamily
  2211.05139}}].

\bibitem{Adshead:2022ovo}
P.~Adshead, P.~Ralegankar and J.~Shelton, \emph{{Dark radiation constraints on
  portal interactions with hidden sectors}},
  \href{https://doi.org/10.1088/1475-7516/2022/09/056}{\emph{JCAP} {\bfseries
  09} (2022) 056} [\href{https://arxiv.org/abs/2206.13530}{{\ttfamily
  2206.13530}}].

\bibitem{McDermott:2010pa}
S.D.~McDermott, H.-B.~Yu and K.M.~Zurek, \emph{{Turning off the Lights: How
  Dark is Dark Matter?}},
  \href{https://doi.org/10.1103/PhysRevD.83.063509}{\emph{Phys. Rev. D}
  {\bfseries 83} (2011) 063509}
  [\href{https://arxiv.org/abs/1011.2907}{{\ttfamily 1011.2907}}].

\bibitem{Dvorkin:2013cea}
C.~Dvorkin, K.~Blum and M.~Kamionkowski, \emph{{Constraining Dark Matter-Baryon
  Scattering with Linear Cosmology}},
  \href{https://doi.org/10.1103/PhysRevD.89.023519}{\emph{Phys. Rev. D}
  {\bfseries 89} (2014) 023519}
  [\href{https://arxiv.org/abs/1311.2937}{{\ttfamily 1311.2937}}].

\bibitem{Kelly:2018brz}
K.J.~Kelly and Y.-D.~Tsai, \emph{{Proton fixed-target scintillation experiment
  to search for millicharged dark matter}},
  \href{https://doi.org/10.1103/PhysRevD.100.015043}{\emph{Phys. Rev. D}
  {\bfseries 100} (2019) 015043}
  [\href{https://arxiv.org/abs/1812.03998}{{\ttfamily 1812.03998}}].

\bibitem{Ball:2016zrp}
A.~Ball et~al., \emph{{A Letter of Intent to Install a milli-charged Particle
  Detector at LHC P5}},  \href{https://arxiv.org/abs/1607.04669}{{\ttfamily
  1607.04669}}.

\bibitem{Planck:2018vyg}
{\scshape Planck} collaboration, \emph{{Planck 2018 results. VI. Cosmological
  parameters}},
  \href{https://doi.org/10.1051/0004-6361/201833910}{\emph{Astron. Astrophys.}
  {\bfseries 641} (2020) A6}
  [\href{https://arxiv.org/abs/1807.06209}{{\ttfamily 1807.06209}}].

\bibitem{Lee:2015zqz}
H.M.~Lee, D.~Kim, K.~Kong and S.C.~Park, \emph{{Diboson Excesses Demystified in
  Effective Field Theory Approach}},
  \href{https://doi.org/10.1007/JHEP11(2015)150}{\emph{JHEP} {\bfseries 11}
  (2015) 150} [\href{https://arxiv.org/abs/1507.06312}{{\ttfamily
  1507.06312}}].

\bibitem{Gurian:2021qhk}
J.~Gurian, D.~Jeong, M.~Ryan and S.~Shandera, \emph{{Molecular Chemistry for
  Dark Matter II: Recombination, Molecule Formation, and Halo Mass Function in
  Atomic Dark Matter}},
  \href{https://doi.org/10.3847/1538-4357/ac75e4}{\emph{Astrophys. J.}
  {\bfseries 934} (2022) 121}
  [\href{https://arxiv.org/abs/2110.11964}{{\ttfamily 2110.11964}}].

\bibitem{Kaplan:2009de}
D.E.~Kaplan, G.Z.~Krnjaic, K.R.~Rehermann and C.M.~Wells, \emph{{Atomic Dark
  Matter}}, \href{https://doi.org/10.1088/1475-7516/2010/05/021}{\emph{JCAP}
  {\bfseries 05} (2010) 021} [\href{https://arxiv.org/abs/0909.0753}{{\ttfamily
  0909.0753}}].

\bibitem{Fan:2013yva}
J.~Fan, A.~Katz, L.~Randall and M.~Reece, \emph{{Double-Disk Dark Matter}},
  \href{https://doi.org/10.1016/j.dark.2013.07.001}{\emph{Phys. Dark Univ.}
  {\bfseries 2} (2013) 139} [\href{https://arxiv.org/abs/1303.1521}{{\ttfamily
  1303.1521}}].

\bibitem{Cyr-Racine:2013fsa}
F.-Y.~Cyr-Racine, R.~de~Putter, A.~Raccanelli and K.~Sigurdson,
  \emph{{Constraints on Large-Scale Dark Acoustic Oscillations from
  Cosmology}}, \href{https://doi.org/10.1103/PhysRevD.89.063517}{\emph{Phys.
  Rev. D} {\bfseries 89} (2014) 063517}
  [\href{https://arxiv.org/abs/1310.3278}{{\ttfamily 1310.3278}}].

\bibitem{Balazs:2022tjl}
C.~Bal\'azs et~al., \emph{{Cosmological constraints on decaying axion-like
  particles: a global analysis}},
  \href{https://doi.org/10.1088/1475-7516/2022/12/027}{\emph{JCAP} {\bfseries
  12} (2022) 027} [\href{https://arxiv.org/abs/2205.13549}{{\ttfamily
  2205.13549}}].

\bibitem{Mirizzi:2009iz}
A.~Mirizzi, J.~Redondo and G.~Sigl, \emph{{Microwave Background Constraints on
  Mixing of Photons with Hidden Photons}},
  \href{https://doi.org/10.1088/1475-7516/2009/03/026}{\emph{JCAP} {\bfseries
  03} (2009) 026} [\href{https://arxiv.org/abs/0901.0014}{{\ttfamily
  0901.0014}}].

\bibitem{Caputo:2020bdy}
A.~Caputo, H.~Liu, S.~Mishra-Sharma and J.T.~Ruderman, \emph{{Dark Photon
  Oscillations in Our Inhomogeneous Universe}},
  \href{https://doi.org/10.1103/PhysRevLett.125.221303}{\emph{Phys. Rev. Lett.}
  {\bfseries 125} (2020) 221303}
  [\href{https://arxiv.org/abs/2002.05165}{{\ttfamily 2002.05165}}].

\bibitem{Caputo:2020rnx}
A.~Caputo, H.~Liu, S.~Mishra-Sharma and J.T.~Ruderman, \emph{{Modeling Dark
  Photon Oscillations in Our Inhomogeneous Universe}},
  \href{https://doi.org/10.1103/PhysRevD.102.103533}{\emph{Phys. Rev. D}
  {\bfseries 102} (2020) 103533}
  [\href{https://arxiv.org/abs/2004.06733}{{\ttfamily 2004.06733}}].

\bibitem{Aramburo-Garcia:2024cbz}
A.~Aramburo-Garcia, K.~Bondarenko, A.~Boyarsky, P.~Kashko, J.~Pradler,
  A.~Sokolenko et~al., \emph{{Dark photon constraints from CMB temperature
  anisotropies}},  \href{https://arxiv.org/abs/2405.05104}{{\ttfamily
  2405.05104}}.

\bibitem{Kogut:2019vqh}
A.~Kogut, M.H.~Abitbol, J.~Chluba, J.~Delabrouille, D.~Fixsen, J.C.~Hill
  et~al., \emph{{CMB Spectral Distortions: Status and Prospects}},
  \href{https://arxiv.org/abs/1907.13195}{{\ttfamily 1907.13195}}.

\bibitem{Li:2008tma}
M.~Li and X.~Zhang, \emph{{Cosmological CPT violating effect on CMB
  polarization}}, \href{https://doi.org/10.1103/PhysRevD.78.103516}{\emph{Phys.
  Rev. D} {\bfseries 78} (2008) 103516}
  [\href{https://arxiv.org/abs/0810.0403}{{\ttfamily 0810.0403}}].

\bibitem{Gluscevic_2012}
V.~Gluscevic, D.~Hanson, M.~Kamionkowski and C.M.~Hirata, \emph{First {CMB}
  constraints on direction-dependent cosmological birefringence from {WMAP}-7},
  \href{https://doi.org/10.1103/physrevd.86.103529}{\emph{Physical Review D}
  {\bfseries 86} (2012) }.

\bibitem{Mei:2014iaa}
H.-H.~Mei, W.-T.~Ni, W.-P.~Pan, L.~Xu and S.~di~Serego~Alighieri, \emph{{New
  constraints on cosmic polarization rotation from the ACTPol cosmic microwave
  background B-Mode polarization observation and the BICEP2 constraint
  update}}, \href{https://doi.org/10.1088/0004-637X/805/2/107}{\emph{Astrophys.
  J.} {\bfseries 805} (2015) 107}
  [\href{https://arxiv.org/abs/1412.8569}{{\ttfamily 1412.8569}}].

\bibitem{Pan:2016vai}
W.P.~Pan, S.~di~Serego~Alighieri, W.T.~Ni and L.~Xu, \emph{{New Constraints On
  Cosmic Polarization Rotation Including SPTpol B-mode Polarization
  Observations}},  in \emph{{2nd LeCosPA Symposium}: {Everything about Gravity,
  Celebrating the Centenary of Einstein's General Relativity}}, 3, 2016,
  \href{https://doi.org/10.1142/9789813203952_0046}{DOI}
  [\href{https://arxiv.org/abs/1603.08193}{{\ttfamily 1603.08193}}].

\bibitem{Contreras:2017sgi}
D.~Contreras, P.~Boubel and D.~Scott, \emph{{Constraints on direction-dependent
  cosmic birefringence from Planck polarization data}},
  \href{https://doi.org/10.1088/1475-7516/2017/12/046}{\emph{JCAP} {\bfseries
  12} (2017) 046} [\href{https://arxiv.org/abs/1705.06387}{{\ttfamily
  1705.06387}}].

\bibitem{Gruppuso:2020kfy}
A.~Gruppuso, D.~Molinari, P.~Natoli and L.~Pagano, \emph{{Planck 2018
  constraints on anisotropic birefringence and its cross-correlation with CMB
  anisotropy}},
  \href{https://doi.org/10.1088/1475-7516/2020/11/066}{\emph{JCAP} {\bfseries
  11} (2020) 066} [\href{https://arxiv.org/abs/2008.10334}{{\ttfamily
  2008.10334}}].

\bibitem{SPT:2020cxx}
{\scshape SPT} collaboration, \emph{{Searching for Anisotropic Cosmic
  Birefringence with Polarization Data from SPTpol}},
  \href{https://doi.org/10.1103/PhysRevD.102.083504}{\emph{Phys. Rev. D}
  {\bfseries 102} (2020) 083504}
  [\href{https://arxiv.org/abs/2006.08061}{{\ttfamily 2006.08061}}].

\bibitem{Namikawa:2020ffr}
T.~Namikawa et~al., \emph{{Atacama Cosmology Telescope: Constraints on cosmic
  birefringence}},
  \href{https://doi.org/10.1103/PhysRevD.101.083527}{\emph{Phys. Rev. D}
  {\bfseries 101} (2020) 083527}
  [\href{https://arxiv.org/abs/2001.10465}{{\ttfamily 2001.10465}}].

\bibitem{Bortolami:2022whx}
M.~Bortolami, M.~Billi, A.~Gruppuso, P.~Natoli and L.~Pagano, \emph{{Planck
  constraints on cross-correlations between anisotropic cosmic birefringence
  and CMB polarization}},
  \href{https://doi.org/10.1088/1475-7516/2022/09/075}{\emph{JCAP} {\bfseries
  09} (2022) 075} [\href{https://arxiv.org/abs/2206.01635}{{\ttfamily
  2206.01635}}].

\bibitem{Zagatti:2024jxm}
G.~Zagatti, M.~Bortolami, A.~Gruppuso, P.~Natoli, L.~Pagano and G.~Fabbian,
  \emph{{Planck constraints on cosmic birefringence and its cross-correlation
  with the CMB}},
  \href{https://doi.org/10.1088/1475-7516/2024/05/034}{\emph{JCAP} {\bfseries
  05} (2024) 034} [\href{https://arxiv.org/abs/2401.11973}{{\ttfamily
  2401.11973}}].

\bibitem{Finelli:2008jv}
F.~Finelli and M.~Galaverni, \emph{{Rotation of Linear Polarization Plane and
  Circular Polarization from Cosmological Pseudo-Scalar Fields}},
  \href{https://doi.org/10.1103/PhysRevD.79.063002}{\emph{Phys. Rev. D}
  {\bfseries 79} (2009) 063002}
  [\href{https://arxiv.org/abs/0802.4210}{{\ttfamily 0802.4210}}].

\bibitem{Alexander:2008fp}
S.~Alexander, J.~Ochoa and A.~Kosowsky, \emph{{Generation of Circular
  Polarization of the Cosmic Microwave Background}},
  \href{https://doi.org/10.1103/PhysRevD.79.063524}{\emph{Phys. Rev. D}
  {\bfseries 79} (2009) 063524}
  [\href{https://arxiv.org/abs/0810.2355}{{\ttfamily 0810.2355}}].

\bibitem{SPIDER:2017kwu}
{\scshape SPIDER} collaboration, \emph{{A New Limit on CMB Circular
  Polarization from SPIDER}},
  \href{https://doi.org/10.3847/1538-4357/aa7cfd}{\emph{Astrophys. J.}
  {\bfseries 844} (2017) 151}
  [\href{https://arxiv.org/abs/1704.00215}{{\ttfamily 1704.00215}}].

\bibitem{Padilla:2019dhz}
I.L.~Padilla et~al., \emph{{Two-year Cosmology Large Angular Scale Surveyor
  (CLASS) Observations: A Measurement of Circular Polarization at 40 GHz}},
  \href{https://arxiv.org/abs/1911.00391}{{\ttfamily 1911.00391}}.

\bibitem{Landau:1975pou}
L.D.~Landau and E.M.~Lifschits, \emph{{The Classical Theory of Fields}},
  vol.~Volume 2 of \emph{Course of Theoretical Physics}, Pergamon Press, Oxford
  (1975).

\end{thebibliography}\endgroup

\end{document}